\documentclass[iop,apjl]{emulateapj}

\usepackage{url}
\usepackage{natbib}
\bibpunct{(}{)}{,}{a}{}{;}
\usepackage{amssymb}
\usepackage{amsmath}
\newcommand{\myemail}{lkristensen@cfa.harvard.edu}

\shorttitle{Methanol as a Signature of the Low-Mass End of the IMF}
\shortauthors{Kristensen \& Bergin}

\begin{document}

\title{Tracing Embedded Stellar Populations in Clusters and Galaxies using Molecular Emission:\\ Methanol as a Signature of the Low-Mass End of the IMF}

\author{Lars E. Kristensen}
\affil{Harvard-Smithsonian Center for Astrophysics, 60 Garden Street, Cambridge, MA 02138, USA}
\email{\myemail}
\author{Edwin A. Bergin}
\affil{Department of Astronomy, University of Michigan, 500 Church Street, Ann Arbor, MI 48109, USA}

\begin{abstract}
Most low-mass protostars form in clusters, in particular high-mass clusters; however, how low-mass stars form in high-mass clusters and what the mass distribution is, are still open questions both in our own Galaxy and elsewhere. To access the population of forming embedded low-mass protostars observationally, we propose to use molecular outflows as tracers. Because the outflow emission scales with mass, the effective contrast between low-mass protostars and their high-mass cousins is greatly lowered. In particular, maps of methanol emission at 338.4 GHz ($J$=7$_0$--6$_0$ A$^+$) in low-mass clusters illustrate that this transition is an excellent probe of the low-mass population. We here present a model of a forming cluster where methanol emission is assigned to every embedded low-mass protostar. The resulting model image of methanol emission is compared to recent ALMA observations toward a high-mass cluster and the similarity is striking: the toy model reproduces observations to better than a factor of two and suggests that approximately 50\% of the total flux originates in low-mass outflows. Future fine-tuning of the model will eventually make it a tool for interpreting the embedded low-mass population of distant regions within our own Galaxy and ultimately higher-redshift starburst galaxies, not just for methanol emission but also water and high-$J$ CO. 
\end{abstract}

\keywords{astrochemistry --- ISM: jets and outflows --- line: profiles --- stars: formation --- stars: winds, outflows}

\section{Introduction}

Most stars form in high-mass clusters \citep{lada03}, and the luminous star-forming regions seen in other galaxies are predominantly (embedded) high-mass clusters. In terms of luminosity, the cluster emission is dominated by the most massive O and B stars, and often global star-formation rates are inferred from their spectra at optical and UV wavelengths \citep{kennicutt12}. Yet even the most massive clusters form very few high-mass stars compared to the total number of stars: the inferred star-formation rate thus depends on the mass distribution in the form of an initial mass function (IMF) to account for the contribution from the low-mass population \citep{hao11}. This population is notoriously difficult to access observationally even in Galactic high-mass clusters beyond Orion \citep[$d$ = 415~pc, e.g.,][]{hillenbrand98}, in particular when this population is still forming and is deeply embedded such as in starburst galaxies. 

Nearby low-mass clusters, on the other hand, are readily studied and both the initial stellar mass and the core mass functions are well-constrained \citep[e.g.][]{andre10, bastian10, offner15}. However, it is unclear if the formation conditions, and thereby the resulting core mass function, are the same in high-mass clusters where protostars are exposed to strong radiation fields from high-mass stars. This poses a problem when extrapolating conditions to extragalactic regions where individual embedded protostars are not resolved. Deeply embedded protostars, both low- and high-mass, are bright in molecular emission and star-forming regions are now traced out to redshifts $>$ 6 \citep{riechers13}, with molecular emission in local galaxies resolved into individual massive clusters \citep[e.g.][]{meier05, meier12, meier14}. Thus, there is a need for connecting the nearby, well-resolved low-mass clusters with first the Galactic massive clusters, and then extrapolating these to nearby extragalactic regions and eventually the higher-redshift starburst galaxies. 

One method for probing the embedded low-mass population is through outflow signatures, one of the key signposts of embedded star formation. The advantage of using outflow emission is that \textit{(i)} low-mass outflow emission scales with envelope mass \citep[e.g.][]{wu04, devilliers14}, a relation which may also hold for high-mass protostars, thereby providing a lower contrast compared to the high-mass protostars in the region, and \textit{(ii)} molecular outflows are uniquely associated with the embedded stages of star formation, both low- and high-mass. This combination makes outflows a potentially excellent probe of the low-mass end of the IMF. 

While low-mass outflow properties (mass, outflow force) are typically measured based on spatially resolved low-$J$ CO emission, CO also traces bulk cloud properties and is therefore not an ideal tracer of outflows and their parent population. Instead, molecules such as methanol are more suitable: the methanol abundance jumps by orders of magnitude in molecular outflows due to sputtering \citep[e.g.][]{bachiller98} and in particular the mid-$J$ transitions with $E_{\rm up}$/$k_{\rm B}$$\sim$50 K appear to be ideal from an empirical point of view \citep{kristensen10a}. These transitions are not excited in the cloud where mostly low-$J$ transitions emit ($E_{\rm up}$/$k_{\rm B}$$\sim$10 K) and they are only moderately excited in hot cores where emission from higher-$J$ transitions is more dominant ($E_{\rm up}$/$k_{\rm B}$$\gtrsim$100 K). The $J$=7$_K$--6$_K$ transitions of methanol around 338~GHz are particularly well suited, both because they are good outflow tracers, and because instruments exist to map emission efficiently \citep[e.g.][]{kristensen10a, torstensson11}. 

In this proof-of-concept study, we present a cluster model which simulates low-mass outflows and their emission based on the known methanol emission from single low-mass protostars in the regions NGC1333 and Serpens Main. The outcome of the model is compared to observations of an embedded high-mass cluster as a step toward calibrating the model before employing it for interpretation of extragalactic observations, in particular by assessing what the high-mass outflow and hot-core contribution to the methanol emission is. The model is described in Sect.~2, and the results are presented in Sect.~3. The model is compared to observations in Sect.~4, where also the implications and future steps are discussed.

\section{Model}

The cluster model consists of two parts: (i) a template cluster with stellar spatial, age, and mass distributions and (ii) molecular emission assigned to each embedded protostar in the cluster. Below, these components are described separately. The codes for running the model are publicly available\footnote{\url{https://github.com/egstrom/cluster-in-a-box}, \\DOI:10.5281/zenodo.13184}. An example of a cluster is shown in Fig.~\ref{fig:distribution} along with the final mass distribution. 

\subsection{Cluster characteristics}

\begin{figure}[t!]
\epsscale{1.2}
\plotone{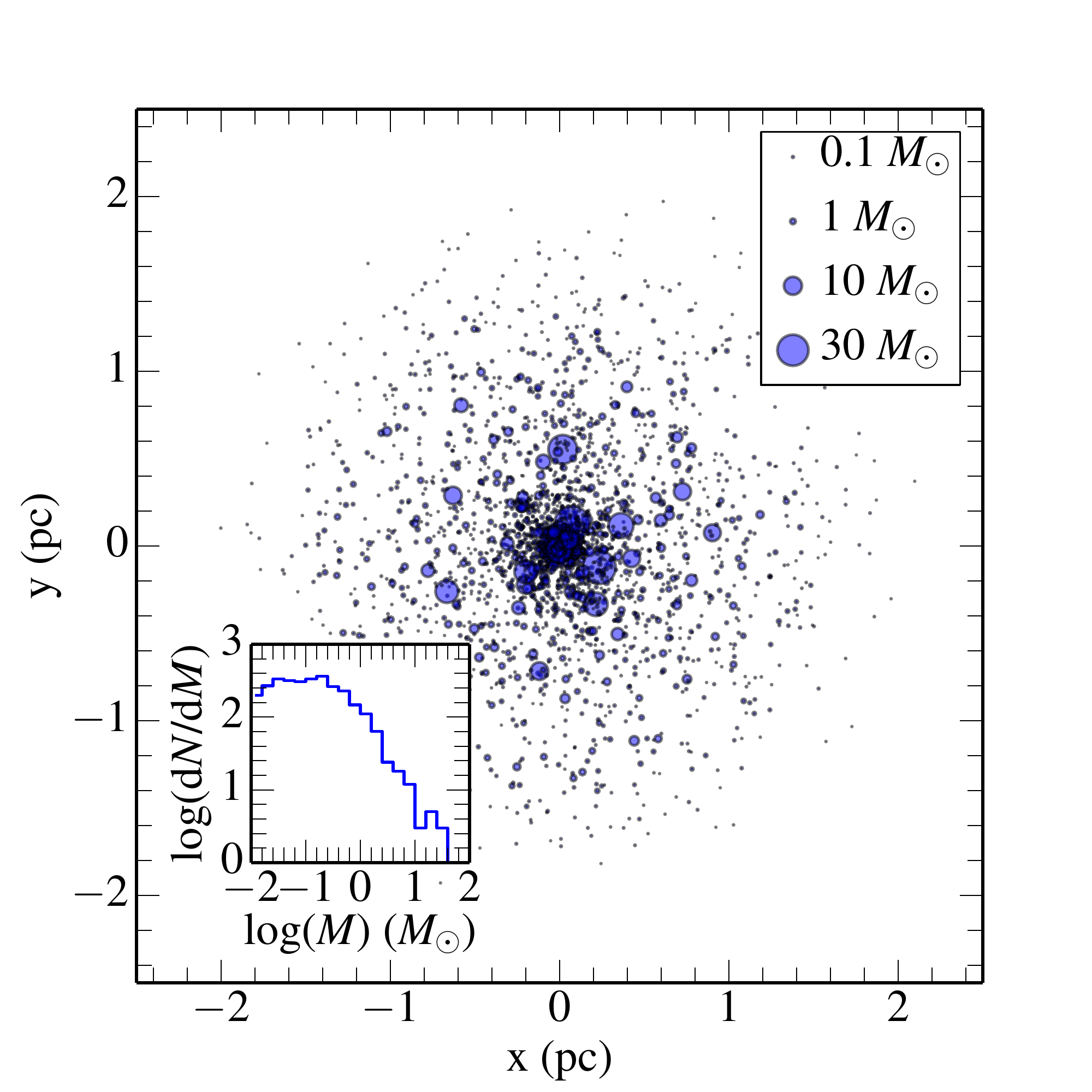}
\caption{Spatial distribution of protostars in the cluster-in-a-box model. This particular cluster contains 3000 stars. The symbol sizes are scaled to the mass of each (proto)star. The inset shows the mass distribution after the low-mass population has been assigned ages and the Class 0/I population assigned new envelope masses. \label{fig:distribution}}
\end{figure}

\textbf{Spatial distribution.} The spatial distribution of protostars builds on the model by \citet{adams14} where the size of the cluster is $R_{\rm max}=R_0 (N/N_0)^\alpha$ and $N$ is the number of stars. This relation works well with $R_0$=1~pc, $N_0$=300 and $\alpha=1/3$ and these are the parameters employed here. The radial probability distribution function is d$P$/d$r$=(3--p)/$R_{\rm max}$ (r/$R_{\rm max}$)$^{2-p}$ where the power-law index $p$ is set to 1.5, again following the example of \citet{adams14}. The angular distribution is assumed to be uniformly random. 

Higher-mass stars are often located closer to the bottom of the gravitational potential well, i.e. closer to the cluster center \citep[e.g.,][]{hillenbrand98}. To simulate this effect in the cluster model, the distance to the center is re-scaled by $M^{-0.1}$ once the spatial distribution is calculated above, a method similar to the method employed by \citet{moeckel09} where the index is 0.15. This small power-law scaling ensures that the most massive members of the cluster are located within 10--20\% of the maximum cluster radius, whereas the lower-mass members are hardly affected. 

\textbf{Age distribution.} The age distribution follows that of the low-mass Perseus star-forming cluster which has an estimated age of $\sim$1~Myr \citep{evans09}. The fraction of protostars in the Class I, II and III stages as well as the ``flat-spectrum'' stage between Class I and II are taken from \citet{evans09}, whereas the Class 0 population is from \citet{sadavoy14}. Perseus is one of the only low-mass clusters where such complete statistics are available. Where outflows only probe the embedded stages, the more evolved stages are revealed through, e.g., infrared observations where good statistics exist for the nearby clouds \citep{evans09}. 

\textbf{Mass distribution.} The initial mass function (IMF) follows that defined in \citep{chabrier03} for young clusters and disk stars and is randomly sampled over the mass interval from 0.01--100~$M_\odot$ (see inset in Fig.~\ref{fig:distribution}). Typically a cluster with 3000 stars contains $\sim$10 high-mass stars ($M>10~M_\odot$) and 1000 brown dwarfs ($M<0.05~M_\odot$). Class 0 cores typically have core masses that are three times higher than the final stellar mass \citep[e.g.][and references therein]{andre10}, and Class I cores have total masses twice as high as the final stellar mass. To account for this excess mass, the total mass of any core in these evolutionary stages is tripled or doubled, respectively. The masses of protostellar systems in the remaining evolutionary classes are typically close to their final mass but since molecular outflows from Class II and III objects are rare and weak, these are not included in the model. 

In this model, the initial number of stars is specified and the total mass of the cluster is subsequently calculated. The stellar mass is from the IMF (see above) and the protostellar masses are obtained by scaling the final stellar mass. The cluster mass is $\sim$10$^3$~$M_\odot$ for a cluster with 3000 stars when all protostellar contributions are summed up. If the star formation efficiency is of the order of 10\% \citep{lada10}, i.e. the total stellar mass divided by the initial cloud mass, the initial cloud would have a mass of the order of $\sim$10$^4$~$M_\odot$ which is typical of observed Galactic molecular clouds. 

\textbf{Outflow angular distribution.} The outflow position angle is randomly set between 0\degr\ and 180\degr. Each outflow lobe is given a random distance from the protostar to simulate both differences in inclination angles and dynamical ages, where the maximum separation corresponds to the maximum separation in the observed data (2$\times$10$^4$~AU; Table~\ref{tab:obs}). 

\subsection{Outflow emission characteristics}

\begin{figure*}[t!]
\epsscale{0.8}
\plotone{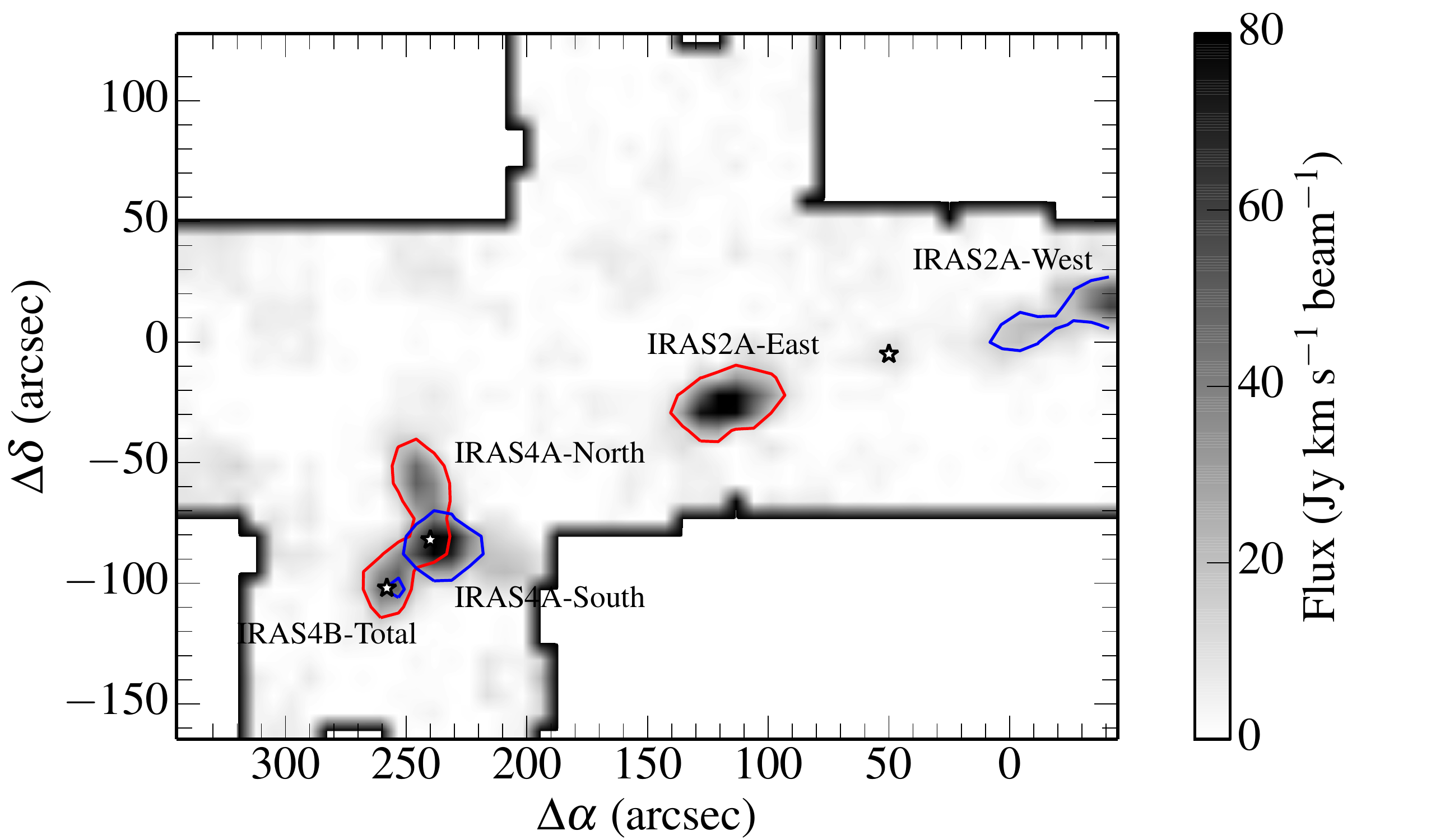}
\epsscale{0.6}
\plotone{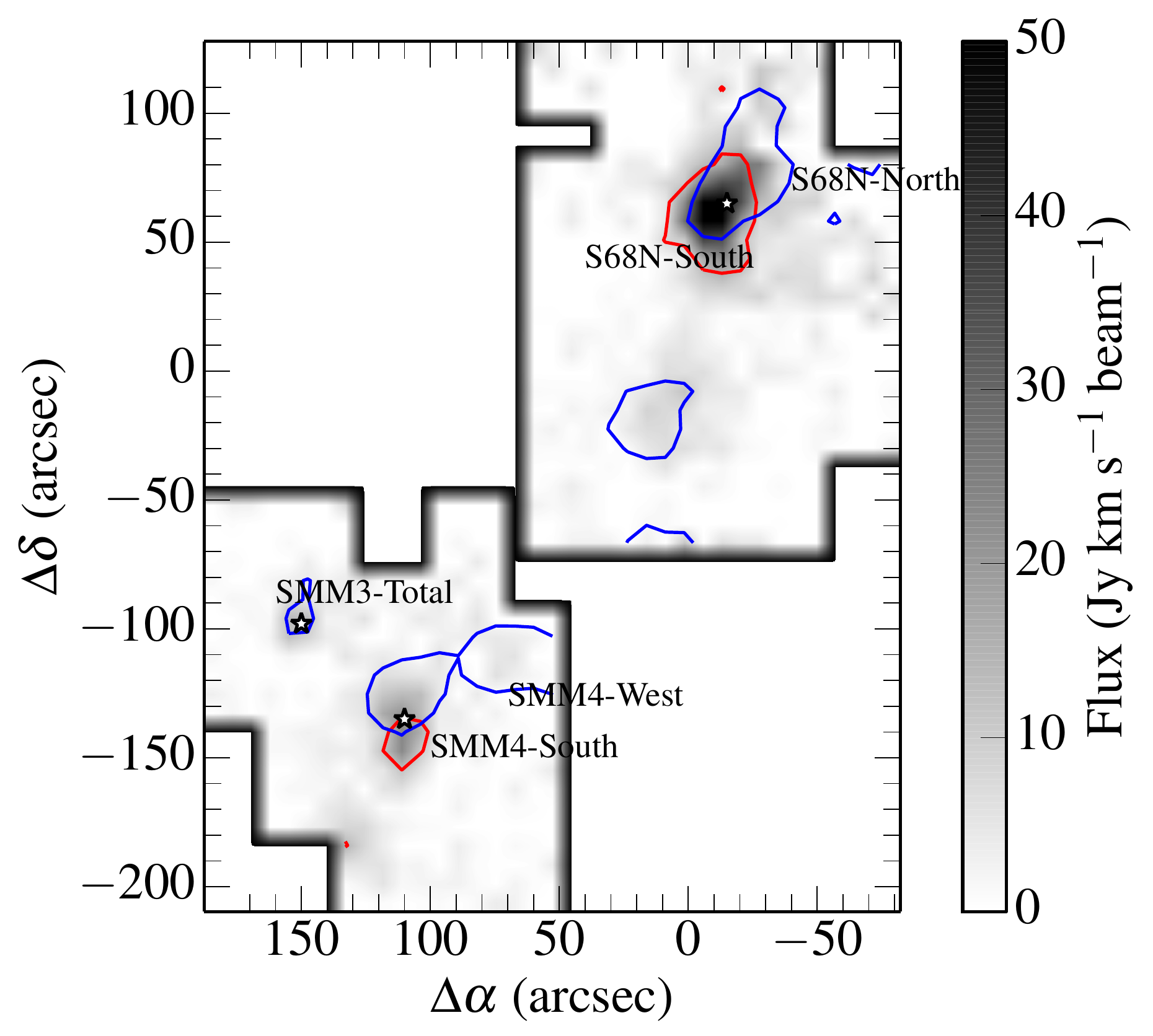}
\caption{Methanol emission toward NGC 1333 (top) and Serpens Main (bottom) obtained at the JCMT. The gray scale is for flux and the contours approximately outline the regions over which outflow emission was integrated per outflow lobe, with red contours for redshifted and blue for blueshifted lobes. In the cases where two lobes are denoted (North/South, East/West), the protostar is located in between them (the protostars are marked with stars). The lobes labeled ``total'' include spatially unresolved emission from both the red- and blue-shifted lobes. The gray lines show the extent of the maps. \label{fig:template}}
\end{figure*}

The outflow template data are obtained from two low-mass star-forming regions, NGC1333 and Serpens Main \citep{kristensen10a}, both mapped with the James Clerk Maxwell Telescope (JCMT) at 338~GHz. Three protostars are contained in the NGC1333 map (IRAS2A, 4A, and 4B) and four in the map of Serpens (SMM1, 3, 4, and S68N). The integrated intensity maps are shown in Fig. \ref{fig:template}. 

In this study, focus is placed on the strongest methanol line around 338 GHz, the 7$_{0}$--6$_{0}$ A$^+$ line at 338.409~GHz ($E_{\rm up}$/$k_{\rm B}$=65~K). The emission maps show empirically that this transition is an excellent outflow tracer, both because of the strong, high-velocity line wings and the spatial distribution \citep[Fig. \ref{fig:template} and][]{kristensen10a}. This is in contrast to lower-excited lines, e.g. the $J$=2--1 transitions around 96~GHz which trace the ambient cloud \citep[e.g.][]{oberg09}, and the higher-excited transitions which typically are associated with hot cores \citep[e.g.][]{menten86}. 

Observations of the $J$ = 7$_0$--6$_0$ A$^+$ transition at 338.409 GHz were obtained with the HARP instrument on the James Clerk Maxwell Telescope (JCMT) on Mauna Kea, Hawaii. HARP is a 16-element array receiver which makes mapping observations very efficient. Both regions were mapped at an angular resolution of 15$''$ over areas of $\sim$ 5$'$ $\times$ 5$'$. The Serpens data are presented in \citet{kristensen10a} where also the data reduction is described in detail. The NGC 1333 data were obtained from project M09AN05 (PI: Kristensen) and were reduced in a similar manner. The integrated intensities are converted from the $T_A^*$ scale to flux by using a conversion factor\footnote{\url{http://docs.jach.hawaii.edu/JCMT/HET/GUIDE/het\_guide/}} of 15.625 Jy K$^{-1}$. 

\begin{figure}[t!]
\epsscale{1.0}
\plotone{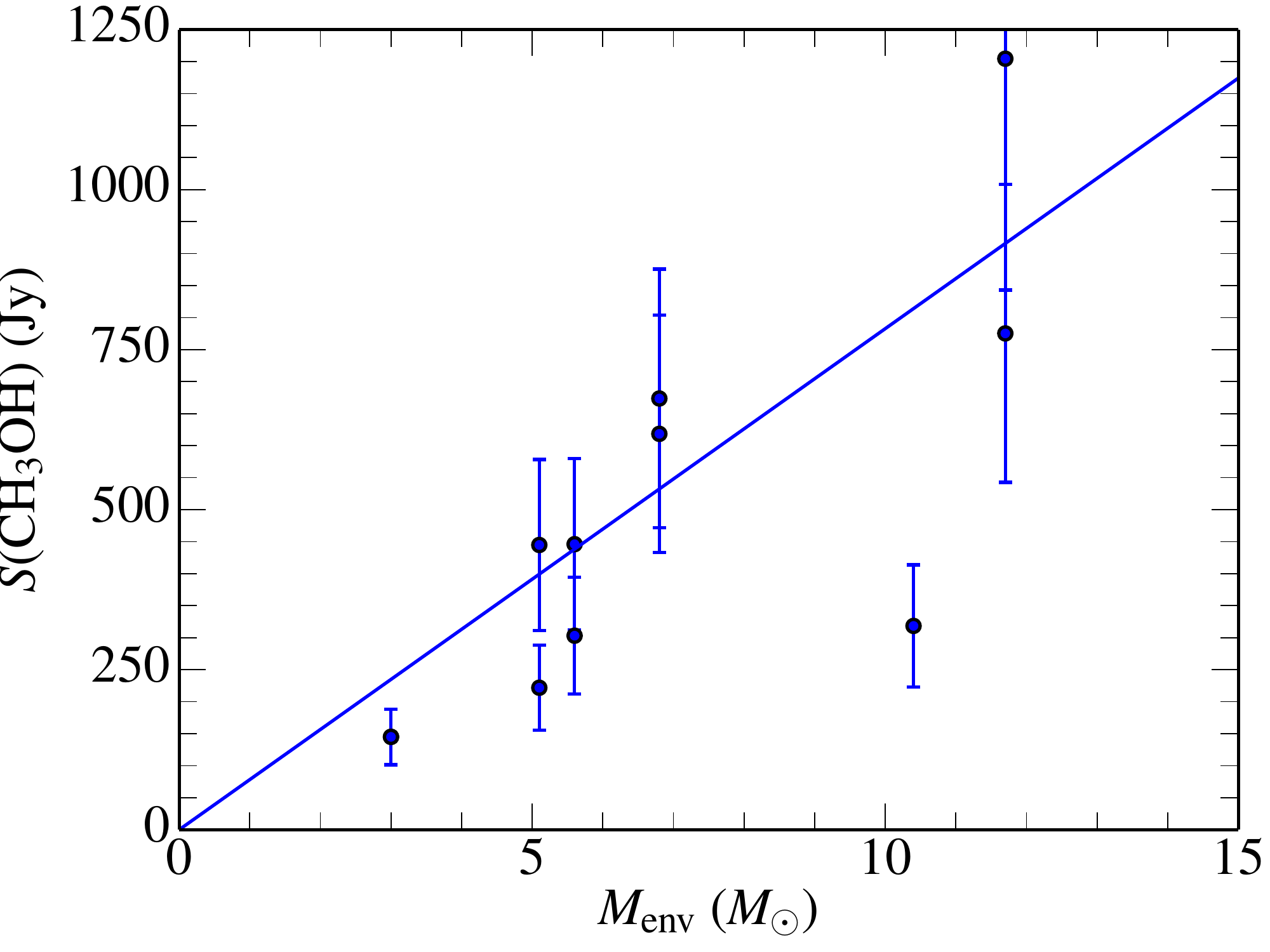}
\caption{Methanol emission versus envelope mass, $M_{\rm env}$, for the sources in NGC1333 and Serpens. The error bars are for the 30\% calibration uncertainty, and the line shows the best-fit proportionality.  \label{fig:ch3oh_corr}}
\end{figure}

Outflow emission from this transition correlates with envelope mass (Fig. \ref{fig:ch3oh_corr}). To test the strength of the correlation, a $t$ test is performed and at the 90\% confidence level there is a direct proportionality between envelope mass and intensity. Because of the low number of data points, the proportionality was not tested for higher levels of significance. The proportionality is $S$ = (78$\pm$33) $M_{\rm env}$ Jy\,km\,s$^{-1}$\,$M_\odot^{-1}$ where the uncertainty is 1$\sigma$. 

To estimate the size of each outflow lobe in the model image, the sizes of the observed outflow lobes are first measured. This is done by measuring the area where the intensity is $>$10\% of the peak, then deconvolving this area with the JCMT beam. Typical deconvolved radii are 1000--3000 AU, with a mean of 2000 AU. At a distance of 3 kpc, this radius corresponds to 0\farcs8 and so is resolvable with ALMA or the SMA, the latter in its very extended configuration. The emission distribution is assumed to be Gaussian for each lobe. The amplitude of the Gaussian is set such that the area matches the observed flux. All observed parameters are listed in Table \ref{tab:obs}. 

\subsection{Outflow emission assignment}

Each low-mass protostellar object is assigned outflow emission based on a scaling of the observed emission with modeled envelope mass; that is, for a given model envelope mass, each simulated outflow lobe is given a total integrated flux corresponding to the scaling of the observed fluxes. 

The outflow activity of Class I sources is typically about an order of magnitude lower than for Class 0 sources \citep{bontemps96}. This lower activity is included in the model by scaling down the emission by a factor of ten for each Class I outflow. Finally, the map is convolved with the desired beam. Note that no radiative transfer is done explicitly, and that all emission is implicitly assumed to be optically thin \citep{bachiller98}. 

\begin{deluxetable}{l c c c c}[th!]
\tabletypesize{\scriptsize}
\tablecaption{Methanol observational parameters. \label{tab:obs}}
\tablewidth{0pt}
\tablehead{\colhead{Source} & \colhead{$M_{\rm env}$\tablenotemark{a}} & \colhead{$L_{\rm bol}\tablenotemark{a}$} & \colhead{$\int\int T_{\rm MB}$ d$v$ d$\Omega$} & \colhead{$\Omega$\tablenotemark{b}} \\
\colhead{} & \colhead{($M_\odot$)} & \colhead{($L_\odot$)} & \colhead{(K km s$^{-1}$)} & \colhead{($n_{\rm pix}$)}
}
\startdata
IRAS2A - East 		& \phantom{1}5.1 	& 35.7 			& 186 			& 23 \\
IRAS2A - West 		& \phantom{1}5.1 	& 35.7 			& 109 			& 27 \\
IRAS4A - North 	& \phantom{1}5.6 	& \phantom{1}9.1 	& \phantom{1}88 	& 16 \\
IRAS4A - South 	& \phantom{1}5.6 	& \phantom{1}9.1 	& 170 			& 21 \\
IRAS4B - Total 		& \phantom{1}3.0 	& \phantom{1}4.4 	& \phantom{1}34 	& 13 \\
SMM4 - South 		& \phantom{1}6.8 	& \phantom{1}6.2 	& \phantom{1}76 	& 21 \\
SMM4 - West 		& \phantom{1}6.8 	& \phantom{1}6.2 	& \phantom{1}59 	& 15 \\
SMM3 - Total 		& 10.4 			& 16.6 			& \phantom{1}22 	& 12 \\
S68N - South 		& 11.7 			& 11.7 			& 148 			& 21 \\
S68N - North 		& 11.7 			& 11.7 			& 136 			& 30
\enddata
\tablenotetext{a}{From \citet{kristensen12}, except for S68N which is from \citet{kristensen10a}. The parameters for the Serpens sources have been scaled to a distance of 415 pc \citep{dzib10}. }
\tablenotetext{b}{Number of 7\farcs5 $\times$ 7\farcs5 pixels.}
\end{deluxetable}

\section{Results}

\begin{figure}[t!]
\epsscale{1.0}
\plotone{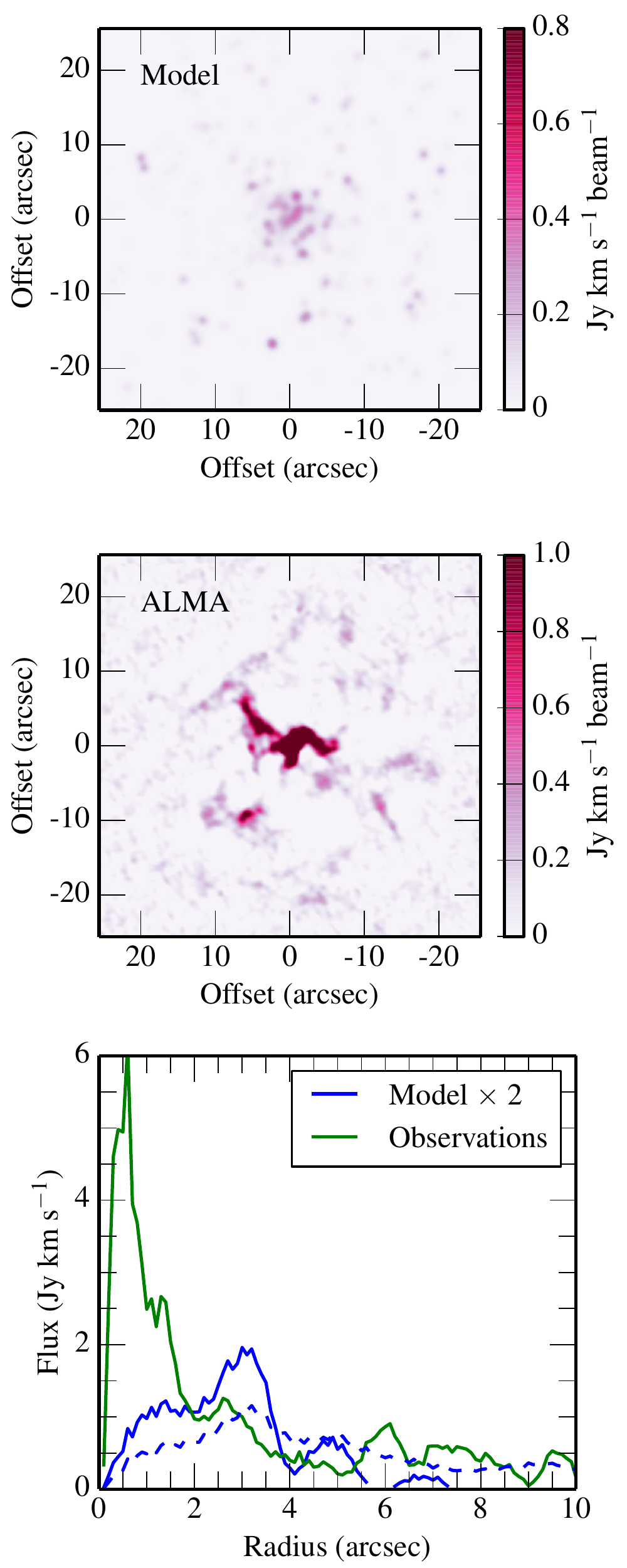}
\caption{\textit{Top:} Integrated intensity map of methanol emission at 338.4 GHz. The image was convolved with a 0\farcs5 beam. The color scale is shown on the right. \textit{Center:} ALMA image of the 338.4 GHz emission toward IRAS16547--4247 from \citet{higuchi14}. \textit{Bottom:} Radially averaged flux from the center of the cluster (blue), multiplied by 2. Also shown is the radially averaged flux profile of the ALMA image (green). Only fluxes $>$3$\sigma$ are included in these profiles. The dashed line shows the radial model profile without $\sigma$-filtering and without scaling. 
\label{fig:image}}
\end{figure}

For a cluster with 3000 protostars and an age distribution of the low-mass population corresponding to that of Perseus, i.e. 1 Myr, the resulting methanol emission map is shown in Fig. \ref{fig:image}. The image has been convolved with a beam of 0\farcs5 and the cluster is assumed to be at a distance of 3 kpc. 

Almost all emission is located in a central condensation where individual outflow lobes are marginally resolved as little clumps (the FWHM of each outflow lobe is 0\farcs8; see above). The maximum separation between outflow lobes is 2$\times$10$^4$ AU, or $\sim$ 7$''$ at a distance of 3 kpc; since each outflow lobe is treated separately as 2D Gaussian functions, two lobes will only be identified as originating in the same protostar if the lobes are close, $\lesssim$2$''$. For a larger beam, such as $\sim$ 2$''$ which is easily achievable with the SubMillimeter Array (SMA), the outflows would have been unresolved. 

The peak intensity is 0.4 Jy km s$^{-1}$ beam$^{-1}$ in the model image. Figure \ref{fig:image} shows the radial flux profile. The profile peaks 3$''$ from the center and remains above 0.5 Jy\,km\,s$^{-1}$ to radii $>$10$''$ (Fig. \ref{fig:image}, dashed line).

\section{Discussion}

\subsection{Comparison to ALMA observations of IRAS16547--4247}

The high-mass star-forming region IRAS16547--4247 was recently observed with ALMA \citep{higuchi14}. The region is located at a distance of 2.9 kpc and was observed at an angular resolution of $\sim$ 0\farcs5 (1500 AU). The data were downloaded from the Science Archive. The pipeline-calibrated visibilities were were first continuum-subtracted using line-free channels in the $uv$ spectrum and then re-imaged; two changes were made compared to the pipelined image presented in \citet{higuchi14}: first, the visibilities were averaged over a spectral region corresponding to velocities from --44 to --20 km\,s$^{-1}$ prior to imaging; second, the clean mask was iteratively placed so that cleaning started at the peak intensity and expanded from there. Both were done to increase image fidelity in the final restored image, with the added bonus of a decrease in noise level of a factor of 2 to 0.07 Jy\,km\,s$^{-1}$\,beam$^{-1}$. The ALMA data are shown in Fig. \ref{fig:image}, where they are not corrected for the shape of the primary beam. For the more quantitative analysis, see below, primary-beam-corrected data are used. The same primary beam is taken into account when analyzing the simulated data. 

The emission is spatially resolved into a central condensation, and a larger number of sub-structures. The central source drives a large-scale CO outflow and part of the methanol emission is clearly associated to the high-mass outflow \citep{higuchi14}. The more wide-spread diffuse methanol emission is likely connected to outflow activity based on the large line-width ($\gtrsim$10~km\,s$^{-1}$). Methanol is only expected to thermally evaporate at $T$$>$100~K \citep{brown07}, and such a region would have a radius of $<$1$''$ \citep{bisschop07}. Thus it appears likely that the spatially extended emission is caused by other, perhaps low-mass, outflow sources as also argued by \citep{higuchi14}. Below, the standard model results are compared to these observations without fine-tuning of the input parameters. 

To compare the model and observations in a quantitative manner, both images are radially averaged from the center of the cluster including only fluxes $>$3$\sigma$. The same observationally set rms cut-off was also applied to the model image. The resulting radial profiles are shown in Fig. \ref{fig:image}. The signal at distances $>$8$''$ may be diluted by false-positives at the 0.1--0.2~Jy\,km\,s$^{-1}$ level. At distances below $\sim$2$''$, fluxes are dominated by the high-mass source (outflow and hot core), and at larger distances the model profile is 30--50\% weaker than the observed profile. If the rms cut-off is not applied to the model image, the radial profile overlaps with observations to $>$10$''$ without a scaling. 

Assuming that the emission at distances $>$ 2$''$ is dominated by the low-mass population, the total flux fraction of the low-mass outflows from the entire cluster can be estimated: approximately 50\% of the cluster flux is coming from the low-mass outflows, the rest is from a combination of the molecular hot core and the high-mass outflow(s). Disentangling these contributions is beyond the scope of this work. 

To obtain the factor of two increase in the model fluxes, several possibilities exist: 
\begin{itemize}
\item \textit{Age.} If the cluster is younger than 1 Myr, the relative number of Class 0 and I sources increase and so will the number low-mass outflows. The model presented in Fig. \ref{fig:image} has $\sim$150 and 400 Class 0 and I sources, respectively, spread over a radius of 150$''$ (but of course centrally peaked). Doubling these numbers would lead to a flux doubling as well. 
\item \textit{IMF.} If the IMF is shifted toward higher masses, which could be the case if the higher-mass stars form before the lower-mass ones, then more methanol emission would be generated per object, because each object would be relatively more massive than in the standard model. 
\item \textit{Size.} The observed cluster may be forming more than 3000 stars in which case more Class 0 and I protostars would need to be included in the model. 
\item \textit{Spatial distribution.} The simulated Class 0 and I protostars are spread randomly over the entire cluster: if the stars form closer to the center, the protostars would be more concentrated and the flux would increase. Indeed, the median distance for a Class 0 protostar to the cluster center is 0.4pc, or 25$''$. 
\end{itemize}
All of these possibilities will be explored quantitatively in the future. However, given that the model is not tuned to reproduce the observations such an overlap is excellent and lends credibility to the model predictions even when using a standard IMF.

\subsection{Future prospects and outlook} 

This proof-of-concept study demonstrates one possible method for gaining access to the embedded low-mass protostellar content of a forming cluster, namely through molecular outflow emission from these low-mass protostars, and thereby potentially constraining the initial (core) mass function. Because outflow emission is directly proportional to the envelope mass, the contrast between low- and high-mass protostars is effectively lowered by several orders of magnitude compared to the bolometric luminosity. 

The success of the model in reproducing part of the morphology of a single observed high-mass embedded cluster paves the way for further fine-tuning of the model, and for including the contribution from high-mass outflows and hot cores explicitly in the model. Two avenues will be pursued in the near future: \textit{(i)} parameter space will be explored, in particular with respect to constraining the initial (core) mass function of low-mass embedded protostars, and \textit{(ii)} other species will be included. As for the latter, \textit{Herschel} observations demonstrate that H$_2$O and high-$J$ CO ($J$$>$16) trace shocks and outflows uniquely, and these are the next species and transitions to be included. Although no new observations of H$_2$O will be done for the foreseeable future, \textit{Herschel} already observed hundreds of Galactic star-forming regions, both low and high mass, thus providing enough data to calibrate the model. Furthermore, ALMA will easily be able to observe H$_2$O emission in redshifted galaxies, and models of this type carry the promise of providing constraints on how low-mass stars have formed and how the IMF has changed throughout cosmological times. 

\acknowledgments
We thank the referee for thoughtful and constructive comments that helped to improve this manuscript.


\begin{thebibliography}{}
\expandafter\ifx\csname natexlab\endcsname\relax\def\natexlab#1{#1}\fi

\bibitem[{{Adams} {et~al.}(2014){Adams}, {Fatuzzo}, \& {Holden}}]{adams14}
{Adams}, F.~C., {Fatuzzo}, M., \& {Holden}, L. 2014, \apj, 789, 86

\bibitem[{{Andr{\'e}} {et~al.}(2010){Andr{\'e}}, {Men'shchikov}, {Bontemps},
  {K{\"o}nyves}, {Motte}, {Schneider}, {Didelon}, {Minier}, {Saraceno},
  {Ward-Thompson}, {di Francesco}, {White}, {Molinari}, {Testi}, {Abergel},
  {Griffin}, {Henning}, {Royer}, {Mer{\'{\i}}n}, {Vavrek}, {Attard},
  {Arzoumanian}, {Wilson}, {Ade}, {Aussel}, {Baluteau}, {Benedettini},
  {Bernard}, {Blommaert}, {Cambr{\'e}sy}, {Cox}, {di Giorgio}, {Hargrave},
  {Hennemann}, {Huang}, {Kirk}, {Krause}, {Launhardt}, {Leeks}, {Le Pennec},
  {Li}, {Martin}, {Maury}, {Olofsson}, {Omont}, {Peretto}, {Pezzuto}, {Prusti},
  {Roussel}, {Russeil}, {Sauvage}, {Sibthorpe}, {Sicilia-Aguilar}, {Spinoglio},
  {Waelkens}, {Woodcraft}, \& {Zavagno}}]{andre10}
{Andr{\'e}}, P., {Men'shchikov}, A., {Bontemps}, S., {et~al.} 2010, \aap, 518,
  L102

\bibitem[{Bachiller} {et~al.}(1998)]{bachiller98} Bachiller, R., Codella, C., Colomer, F., Liechti, S., \& Walmsley, C.~M.\ 1998, \aap, 335, 266 

\bibitem[{{Bastian} {et~al.}(2010){Bastian}, {Covey}, \& {Meyer}}]{bastian10}
{Bastian}, N., {Covey}, K.~R., \& {Meyer}, M.~R. 2010, \araa, 48, 339

\bibitem[{Bisschop} {et~al.}(2007)]{bisschop07} Bisschop, S.~E., J{\o}rgensen, J.~K., van Dishoeck, E.~F., \& de Wachter, E.~B.~M.\ 2007, \aap, 465, 913 

\bibitem[{{Bontemps} {et~al.}(1996){Bontemps}, {Andr{\'e}}, {Terebey}, \&
  {Cabrit}}]{bontemps96}
{Bontemps}, S., {Andr{\'e}}, P., {Terebey}, S., \& {Cabrit}, S. 1996, \aap,
  311, 858

\bibitem[{Brown} \& {Bolina}(2007)]{brown07} Brown, W.~A., \& Bolina, A.~S.\ 2007, \mnras, 374, 1006 

\bibitem[{{Chabrier}(2003)}]{chabrier03}
{Chabrier}, G. 2003, \pasp, 115, 763

\bibitem[{{Dzib} {et~al.}(2010){Dzib}, {Loinard}, {Mioduszewski}, {Boden},
  {Rodr{\'{\i}}guez}, \& {Torres}}]{dzib10}
{Dzib}, S., {Loinard}, L., {Mioduszewski}, A.~J., {et~al.} 2010, \apj, 718, 610

\bibitem[{{Evans} {et~al.}(2009){Evans}, {Dunham}, {J{\o}rgensen}, {Enoch},
  {Mer{\'{\i}}n}, {van Dishoeck}, {Alcal{\'a}}, {Myers}, {Stapelfeldt},
  {Huard}, {Allen}, {Harvey}, {van Kempen}, {Blake}, {Koerner}, {Mundy},
  {Padgett}, \& {Sargent}}]{evans09}
{Evans}, N.~J., {Dunham}, M.~M., {J{\o}rgensen}, J.~K., {et~al.} 2009, \apjs,
  181, 321

\bibitem[{{Hao} {et~al.}(2011){Hao}, {Kennicutt}, {Johnson}, {Calzetti},
  {Dale}, \& {Moustakas}}]{hao11}
{Hao}, C.-N., {Kennicutt}, R.~C., {Johnson}, B.~D., {et~al.} 2011, \apj, 741,
  124

\bibitem[{{Higuchi} {et~al.}(2015){Higuchi}, {Saigo}, {Chibueze}, {Sanhueza},
  {Takakuwa}, \& {Garay}}]{higuchi14}
{Higuchi}, A.~E., {Saigo}, K., {Chibueze}, J.~O., {et~al.} 2015, \apjl, 798,
  L33

\bibitem[{{Hillenbrand} \& {Hartmann}(1998)}]{hillenbrand98}
{Hillenbrand}, L.~A., \& {Hartmann}, L.~W. 1998, \apj, 492, 540

\bibitem[{{Kennicutt} \& {Evans}(2012)}]{kennicutt12}
{Kennicutt}, R.~C., \& {Evans}, N.~J. 2012, \araa, 50, 531

\bibitem[{{Kristensen} {et~al.}(2010){Kristensen}, {van Dishoeck}, {van
  Kempen}, {Cuppen}, {Brinch}, {J{\o}rgensen}, \&
  {Hogerheijde}}]{kristensen10a}
{Kristensen}, L.~E., {van Dishoeck}, E.~F., {van Kempen}, T.~A., {et~al.} 2010,
  \aap, 516, A57

\bibitem[{{Kristensen} {et~al.}(2012){Kristensen}, {van Dishoeck}, {Bergin},
  {Visser}, {Y{\i}ld{\i}z}, {San Jose-Garcia}, {J{\o}rgensen}, {Herczeg},
  {Johnstone}, {Wampfler}, {Benz}, {Bruderer}, {Cabrit}, {Caselli}, {Doty},
  {Harsono}, {Herpin}, {Hogerheijde}, {Karska}, {van Kempen}, {Liseau},
  {Nisini}, {Tafalla}, {van der Tak}, \& {Wyrowski}}]{kristensen12}
{Kristensen}, L.~E., {van Dishoeck}, E.~F., {Bergin}, E.~A., {et~al.} 2012,
  \aap, 542, A8

\bibitem[{{Lada} \& {Lada}(2003)}]{lada03}
{Lada}, C.~J., \& {Lada}, E.~A. 2003, \araa, 41, 57

\bibitem[{{Lada} {et~al.}(2010){Lada}, {Lombardi}, \& {Alves}}]{lada10}
{Lada}, C.~J., {Lombardi}, M., \& {Alves}, J.~F. 2010, \apj, 724, 687

\bibitem[{{Meier} \& {Turner}(2005)}]{meier05}
{Meier}, D.~S., \& {Turner}, J.~L. 2005, \apj, 618, 259

\bibitem[{{Meier} \& {Turner}(2012)}]{meier12}
---. 2012, \apj, 755, 104

\bibitem[{{Meier} {et~al.}(2014){Meier}, {Turner}, \& {Beck}}]{meier14}
{Meier}, D.~S., {Turner}, J.~L., \& {Beck}, S.~C. 2014, \apj, 795, 107

\bibitem[{{Menten} {et~al.}(1986){Menten}, {Walmsley}, {Henkel}, {Wilson},
  {Snyder}, {Hollis}, \& {Lovas}}]{menten86}
{Menten}, K.~M., {Walmsley}, C.~M., {Henkel}, C., {et~al.} 1986, \aap, 169, 271

\bibitem[{{Moeckel} \& {Bonnell}(2009)}]{moeckel09}
{Moeckel}, N., \& {Bonnell}, I.~A. 2009, \mnras, 396, 1864

\bibitem[{{{\"O}berg} {et~al.}(2009){{\"O}berg}, {Bottinelli}, \& {van
  Dishoeck}}]{oberg09}
{{\"O}berg}, K.~I., {Bottinelli}, S., \& {van Dishoeck}, E.~F. 2009, \aap, 494,
  L13

\bibitem[{{Offner} {et~al.}(2014){Offner}, {Clark}, {Hennebelle}, {Bastian},
  {Bate}, {Hopkins}, {Moraux}, \& {Whitworth}}]{offner15}
{Offner}, S.~S.~R., {Clark}, P.~C., {Hennebelle}, P., {et~al.} 2014, Protostars
  and Planets VI, 53

\bibitem[{{Riechers} {et~al.}(2013){Riechers}, {Bradford}, {Clements},
  {Dowell}, {P{\'e}rez-Fournon}, {Ivison}, {Bridge}, {Conley}, {Fu}, {Vieira},
  {Wardlow}, {Calanog}, {Cooray}, {Hurley}, {Neri}, {Kamenetzky}, {Aguirre},
  {Altieri}, {Arumugam}, {Benford}, {B{\'e}thermin}, {Bock}, {Burgarella},
  {Cabrera-Lavers}, {Chapman}, {Cox}, {Dunlop}, {Earle}, {Farrah}, {Ferrero},
  {Franceschini}, {Gavazzi}, {Glenn}, {Solares}, {Gurwell}, {Halpern},
  {Hatziminaoglou}, {Hyde}, {Ibar}, {Kov{\'a}cs}, {Krips}, {Lupu}, {Maloney},
  {Martinez-Navajas}, {Matsuhara}, {Murphy}, {Naylor}, {Nguyen}, {Oliver},
  {Omont}, {Page}, {Petitpas}, {Rangwala}, {Roseboom}, {Scott}, {Smith},
  {Staguhn}, {Streblyanska}, {Thomson}, {Valtchanov}, {Viero}, {Wang},
  {Zemcov}, \& {Zmuidzinas}}]{riechers13}
{Riechers}, D.~A., {Bradford}, C.~M., {Clements}, D.~L., {et~al.} 2013, \nat,
  496, 329

\bibitem[{{Sadavoy} {et~al.}(2014){Sadavoy}, {Di Francesco}, {Andr{\'e}},
  {Pezzuto}, {Bernard}, {Maury}, {Men'shchikov}, {Motte}, {Nguy{\^e}n-Lu'o'ng},
  {Schneider}, {Arzoumanian}, {Benedettini}, {Bontemps}, {Elia}, {Hennemann},
  {Hill}, {K{\"o}nyves}, {Louvet}, {Peretto}, {Roy}, \& {White}}]{sadavoy14}
{Sadavoy}, S.~I., {Di Francesco}, J., {Andr{\'e}}, P., {et~al.} 2014, \apjl,
  787, L18

\bibitem[{{Torstensson} {et~al.}(2011){Torstensson}, {van der Tak}, {van
  Langevelde}, {Kristensen}, \& {Vlemmings}}]{torstensson11}
{Torstensson}, K.~J.~E., {van der Tak}, F.~F.~S., {van Langevelde}, H.~J.,
  {Kristensen}, L.~E., \& {Vlemmings}, W.~H.~T. 2011, \aap, 529, A32

\bibitem[{de~Villiers} {et~al.}(2014)]{devilliers14} de Villiers, H.~M., 
Chrysostomou, A., Thompson, M.~A., et al.\ 2014, \mnras, 444, 566 

\bibitem[{Wu} {et~al.}(2004)]{wu04} Wu, Y., Wei, Y., Zhao, M., et al.\ 2004, \aap, 426, 503 

\end{thebibliography}
\end{document}